\documentclass{article}

\usepackage{arxiv}

\usepackage[utf8]{inputenc} % allow utf-8 input
\usepackage[T1]{fontenc}    % use 8-bit T1 fonts
\usepackage{hyperref}       % hyperlinks
\usepackage{url}            % simple URL typesetting
\usepackage{booktabs}       % professional-quality tables
\usepackage{amsfonts}       % blackboard math symbols
\usepackage{nicefrac}       % compact symbols for 1/2, etc.
\usepackage{microtype}      % microtypography
\usepackage{lipsum}         % Can be removed after putting your text content
\usepackage{graphicx}
\usepackage{doi}

\usepackage[numbers,sort&compress]{natbib}

\usepackage{caption}
\usepackage[labelformat=simple]{subcaption}

\usepackage{graphicx}
\usepackage{textcomp}
\usepackage{amsmath}
\usepackage{float}
\usepackage{siunitx}
\usepackage{array}
\usepackage{booktabs}
\usepackage{longtable}

\usepackage[ruled]{algorithm2e}
\usepackage{xcolor}
\SetKwComment{Comment}{$\triangleright$\ }{}

\newcommand\HLBM{HLBM}
\newcommand\Rey{\mbox{\textit{Re}}}

\definecolor{kit-green}{cmyk}{1, 0, 0.6, 0}
\definecolor{kit-blue}{cmyk}{0.8, 0.5, 0, 0}
\definecolor{kit-maygreen}{cmyk}{0.6, 0, 1, 0}
\definecolor{kit-yellow}{cmyk}{0, 0.05, 1, 0}
\definecolor{kit-orange}{cmyk}{0, 0.45, 1, 0}
\definecolor{kit-brown}{cmyk}{0.35, 0.5, 1, 0}
\definecolor{kit-red}{cmyk}{0.25, 1, 1, 0}
\definecolor{kit-purple}{cmyk}{0.25, 1, 0, 0}
\definecolor{kit-cyan}{cmyk}{0.9, 0.05, 0, 0}
\definecolor{kit-black}{cmyk}{0, 0, 0, 1}

\usepackage{cleveref}

\title{A novel particle decomposition scheme to improve parallel performance of fully resolved particulate flow simulations}

\author{
	\href{https://orcid.org/0000-0002-7666-3439}{\includegraphics[scale=0.06]{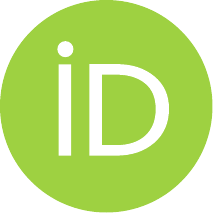}\hspace{1mm}Jan E. Marquardt}\thanks{Corresponding author}\\
	Lattice Boltzmann Research Group \\
	Institute of Mechanical Process Engineering and Mechanics\\
	Karlsruhe Institute of Technology\\
	76131 Karlsruhe, Germany \\
	\texttt{jan.marquardt@kit.edu} \\
	\And
	\href{https://orcid.org/0000-0003-3641-4873}{\includegraphics[scale=0.06]{orcid.pdf}\hspace{1mm}Nicolas Hafen}\\
	Lattice Boltzmann Research Group \\
	Institute of Mechanical Process Engineering and Mechanics \\
	Karlsruhe Institute of Technology \\
	76131 Karlsruhe, Germany \\
	\And
	\href{https://orcid.org/0000-0003-1026-6462}{\includegraphics[scale=0.06]{orcid.pdf}\hspace{1mm}Mathias J. Krause}\\
	Lattice Boltzmann Research Group \\
	Institute for Applied and Numerical Mathematics \\
	Karlsruhe Institute of Technology \\
	76131 Karlsruhe, Germany \\
}

\renewcommand{\headeright}{J.\ E.\ Marquardt \emph{et al.}}
\renewcommand{\shorttitle}{A novel particle decomposition scheme}

\hypersetup{
	pdftitle={SurfaceResolvedParticleDecompositionScheme},
	pdfsubject={},
	pdfauthor={Jan E. Marquardt}
}

\begin{document}
	\maketitle

\begin{abstract}
This study addresses the challenge of simulating realistic particle systems by proposing a novel particle decomposition scheme that improves the parallel performance of surface resolved particle simulations.
Realistic particle systems often involve large numbers of particles and complex particle shapes.
The resulting need to account for shape factors requires the inclusion of even more particles to obtain statistically meaningful results.
However, the computational cost increases with the number of particles, making efficient parallelization crucial.
Therefore, the proposed scheme aims to improve the scalability by optimizing the communication and data management between processors.
Through hindered settling experiments, the applicability and performance of the novel particle decomposition scheme are thoroughly investigated using the homogenized lattice Boltzmann method.
The results show that the proposed method significantly improves the performance, especially in scenarios with a large number of particles, by reducing communication constraints and improving scalability.
This research contributes to the advancement of computational methods for efficient study of complex particle systems and provides valuable insights for future developments in this field.
\end{abstract}

\keywords{
	particle-resolved simulation  \and 
	homogenized lattice Boltzmann method \and 
	partially saturated methods \and 
	hindered settling \and 
	OpenLB.}

\section{Introduction}

In numerous industrial and scientific contexts, the study of particulate systems has emerged as a critical area of research.
Particles have an astonishing variety of complex shapes, making their behavior highly complex and difficult to understand.
Understanding and manipulating the behavior of these complexly shaped particles is vital to improving existing industrial processes and developing new technological advances.
\par
The extensive experimental and numerical research devoted to the study of hindered settling highlights the importance of studying particle collectives. 
Research initially focused on obtaining correlations for the average settling velocity in order to establish relationships between suspension characteristics and settling behavior \cite{Steinour_1944, Richardson_Zaki_1954, Barnea_Mizrahi_1973, Garside_Al-Dibouni_1977, Di_Felice_1995}. 
These early studies provided valuable insights into the collective behavior of particles and laid the foundation for further investigations of dilute \cite{Di_Felice_1999} and concentrated suspensions at moderate Reynolds numbers \cite{Hamid_Molina_Yamamoto_2014, Zaidi_Tsuji_Tanaka_2015}.
Subsequently, the research expanded to include the formation and dynamics of particle clusters during hindered settling~\cite{Willen_Prosperetti_2019, Yao_Criddle_Fringer_2021}. 
Understanding the clustering is crucial as it significantly influences the overall settling behavior and the efficiency of separation processes. 
\par
Despite these extensive research efforts, it is important to note that the above studies have focused solely on simple spherical geometries.
However, in many industrial and scientific applications, particles have non-spherical and usually irregular shapes.
Therefore, several experimental studies consider more complex shapes, such as cubic and brick-like shapes~\cite{Chong_Ratkowsky_Epstein_1979}, fibers~\cite{Jirout_Jiroutova_2022}, and irregular sand grains~\cite{Tomkins_Baldock_Nielsen_2005}.
Current knowledge, however, is limited to explicit shapes, and a deeper understanding, including correlations of the average settling velocity with shape factors, remains a challenge.
Experiments, though valuable, are time-consuming, costly, and inherently difficult to control and adjust specific parameters, such as particle shapes.
In addition, it is difficult to resolve a large number of particles and high particle volume fractions.
\par
Numerical investigations therefore play an essential role in gaining a thorough understanding of such systems.
However, comprehensive models are complex and computationally expensive due to the need to incorporate four-way coupling.
This includes coupling between fluid and particles, between particles themselves, and between particles and walls.
It is essential to accurately capture the complex interactions and dynamics within the system.
In addition, the presence of complex shapes introduces shape factors that affect collective behavior.
Hence, to obtain statistically meaningful results, a larger number of particles is required, which further increases computational demands.
\par
The discrete element method (DEM) is one approach to consider particulate flows.
It is widely used for simulations in many fields \cite{Zhu_Zhou_Yang_Yu_2008} and can represent arbitrary geometries using the glued-sphere technique \cite{Nolan_Kavanagh_1995} or similarly by combining other convex shapes \cite{Rakotonirina_Delenne_Radjai_Wachs_2019}.
However, approximating arbitrary shapes from combinations results in limited accuracy.
To improve the accuracy, the number of segments must be increased significantly, resulting in expensive computations.
In addition, modeling the influence of the surrounding fluid is challenging.
While there have been studies on coupling with a surrounding fluid, these mainly use simple spherical geometries \cite{Qiu_Wu_2014,Sun_Xiao_2016} or model only the drag coefficient without back-coupling from the particles to the fluid \cite{weers2022DevelopmentModelSeparation}.
However, the back-coupling is of enormous importance, especially at high particle volume fractions \cite{andersson2011computational}.
\par
The immersed boundary method (IBM) is another well-known option to consider non-spherical particles, since it resolves surfaces as Lagrange points~\cite{Uhlmann_2005}.
The interaction of these points with the fluid is independent of the fluid grid, resulting in a high level of accuracy.
In addition, a significant advantage of IBM is its ability to be coupled with various fluid solvers, including the finite element method and the lattice Boltzmann method (LBM).
For example, it has been used to study the settling behavior of elliptical particles \cite{AmiriDelouei2022,Karimnejad2018}, which has also been addressed by other LBM-based methods~\cite{Hu2015}.
However, the frequent interpolations between particle and fluid points are computationally expensive.
\par
The partially saturated method (PSM), originally introduced by Noble and Torczynski \cite{Noble_Torczynski_1998} is the most widely used alternative LBM-based method for simulating particles of arbitrary shape.
This preference for PSM is consistent with the growing interest in LBM due to its efficiency and ease of parallelization, driven by the fact that its computationally intensive operations are inherently localized~\cite{Succi_2001}.
Subsequently, various new derivatives have been introduced~\cite{Haussmann_Hafen_Raichle_Trunk_Nirschl_Krause_2020,Rettinger_Rüde_2017}, including the homogenized lattice Boltzmann method (HLBM)~\cite{Krause_Klemens_Henn_Trunk_Nirschl_2017,Trunk_Marquardt_Thaeter_Nirschl_Krause_2018,Trunk_Weckerle_Hafen_Thaeter_Nirschl_Krause_2021}.
The ability to accurately represent a wide range of shapes has been demonstrated in several cases. For example, this capability has been used to derive shape-dependent drag coefficient correlations~\cite{Trunk_Bretl_Thaeter_Nirschl_Dorn_Krause_2021}.
In addition, \HLBM{} has been applied to the simulation of cubic disk-shaped particles in wall flow filters~\cite{Hafen_Dittler_Krause_2022, Hafen_2023, Hafen_Marquardt_Dittler_Krause_2023,Hafen2023c} and another PSM derivative to the filtration of irregular airborne particles~\cite{Li2019}.
The potential applications are further extended by a compatible contact model that is suitable for arbitrary convex shapes~\cite{Marquardt_Römer_Nirschl_Krause_2022}.
Since this feature enables four-way coupled simulations using PSMs that already take advantage of the LBM's parallelization capabilities to improve the particle-fluid coupling performance.
However, the common implementation of PSMs involves communicating all data to each process involved in the computation, which can lead to significant overhead and limit the scalability of simulations, especially at higher particle counts and domain sizes.
\par
In summary, direct numerical simulations are essential for the numerical study of suspensions consisting of arbitrarily shaped particles, and several approaches, such as PSM and HLBM, are available for such simulations.
However, these approaches share a significant computational cost, which poses a challenge in their application to more realistic scenarios involving hundreds or even thousands of particles.
Consequently, there is an urgent need to increase their efficiency.
\par
The aim of the present work is therefore to improve the efficiency of PSMs through the development of a novel particle decomposition scheme that allows for more efficient simulations.
By achieving this objective, future PSM simulations are conductable with larger particle populations, allowing for more accurate and comprehensive investigations of complex-shaped particle systems.
This eventually increases its applicability to real-world problems.
\par
To this end, the remainder of this paper is organized as follows.
Section~\ref{sec:modeling} introduces the models used to consider the fluid and the particles, while Section~\ref{sec:methods} discusses the numerical methods used to solve the model system, followed by the proposed domain decomposition scheme in Section~\ref{sec:decomposition} and its application to hindered settling in Section~\ref{sec:application}.
Finally, Section~\ref{sec:summary} provides a summary and conclusion.

\section{Modeling}
\label{sec:modeling}

\subsection{Fluid}
\label{sec:modeling:fluid}

Fluids are commonly considered to be incompressible.
In this case, the Navier--Stokes equations become
\begin{align}
	\begin{split}
		\frac{\partial \boldsymbol{u}_\text{f}}{\partial t} + \left( \boldsymbol{u}_\text{f} \cdot \nabla \right) \boldsymbol{u}_\text{f} - \nu \Delta \boldsymbol{u}_\text{f} + \frac{1}{\rho_\text{f}} \nabla p &= \boldsymbol{F}_\text{f},
		\\
		\nabla \cdot \boldsymbol{u}_\text{f} &= 0,
	\end{split}
\end{align}
where $p$ is the pressure, $t$ is the time, $\boldsymbol{F}_\text{f}$ is the total of all forces acting on the fluid, and $\boldsymbol{u}_\text{f}$, $\rho_\text{f}$, $\nu$ are the velocity, density, and kinematic viscosity of the fluid.

\subsection{Particle}
\label{sec:modeling:particle}

For the particle component, we use Newton's second law of motion. Therefore, translation is described by
\begin{align}
	\label{eq:particlemotiontrans}
	{m_\text{p}} \frac{\partial \boldsymbol{u}_\text{p}}{\partial t} = \boldsymbol{F}_\text{p}
\end{align}
and rotation by
\begin{equation}
	\label{eq:motionrot}
	\boldsymbol{I}_\text{p} \frac{\partial \boldsymbol{\omega}_\text{p}}{\partial t} +  \boldsymbol{\omega}_\text{p} \times (\boldsymbol{I}_\text{p} \cdot \boldsymbol{\omega}_\text{p}) = \boldsymbol{T}_\text{p}.
\end{equation}
Here, ${m_\text{p}}$, $\boldsymbol{I}_\text{p}$, $\boldsymbol{u}_\text{p}$,  $\boldsymbol{\omega}_\text{p}$ are the mass, moment of inertia, velocity, and angular velocity of the particle.
$\boldsymbol{F}_\text{p}$ and $\boldsymbol{T}_\text{p}$ are the sum of the acting forces and torques affecting the particle motion, which may include the hydrodynamic forces and torques mentioned in Section~\ref{sec:methods:psm} or contact treatment results~\cite{Marquardt_Römer_Nirschl_Krause_2022,Karimnejad2022}.
Above, the subscript $\text{p}$ indicates that the quantities refer to the particle's center of mass.

\section{Numerical Methods}
\label{sec:methods}

\subsection{Lattice Boltzmann Method}
\label{sec:methods:lbm}

In this work, the Navier--Stokes equations for incompressible flows are solved using the lattice Boltzmann method (LBM) \cite{Succi_2001,Krueger2016,sukop2006LatticeBoltzmannModeling}. 
Note that all values in Section~\ref{sec:methods} are given in lattice units, unless explicitly stated otherwise.
\par
LBM has its roots in gas kinetics, which explains why it operates at the mesoscopic level and considers the behavior of particle populations.
Accordingly, particles in this section refer to the fluid particles.
The discrete velocity distribution function $f_i(\boldsymbol{x},t)$ is used to characterize the aforementioned populations at a position $\boldsymbol{x}$ and time $t$.
The index $i$ refers to the corresponding discrete velocity $\boldsymbol{c}_i$, which is given by the selected velocity set.
There are several velocity sets available in the literature \cite{Succi_2001,Krueger2016}.
For the studies in this paper, we choose the D$3$Q$19$, which discretizes the three-dimensional space and contains $19$ discrete velocities
\begin{equation}
	\boldsymbol{c}_i = 
	\begin{cases}
		(0,0,0), &\text{if } i=0 \\
		(\pm 1, 0, 0), (0, \pm 1, 0), (0, 0, \pm 1), &\text{if } i = 1, \ldots, 6 \\
		(\pm 1, \pm 1, 0), (\pm 1, 0, \pm 1), (0, \pm 1, \pm 1), &\text{if } i = 7, \ldots, 18 \\
	\end{cases}.
\end{equation}

The populations are also used to derive macroscopic quantities such as the fluid density $\rho_\text{f}(\boldsymbol{x},t) = \sum_i f_i(\boldsymbol{x},t)$ and velocity $\rho_\text{f} \boldsymbol{u}_\text{f}(\boldsymbol{x},t) = \sum_i\boldsymbol{c}_i f_i(\boldsymbol{x},t)$.
\par
The particle populations' evolution over time is expressed by the lattice Boltzmann equation that is usually divided into a collision and streaming step.
The former reads
\begin{equation}
	\label{eq:collide}
	f_i^*(\boldsymbol{x}, t) = f_i(\boldsymbol{x}, t) + \Omega_i(\boldsymbol{x}, t) + S_i(\boldsymbol{x}, t).
\end{equation}
Here, the post-collision distribution $f_i^*$ is obtained using a collision operator $\Omega_i$ and an optional source term $S_i$.
Furthermore, the propagation step with $\Delta t = \Delta x = 1$ is given by
\begin{equation}
	\label{eq:stream}
	f_i(\boldsymbol{x} + \boldsymbol{c}_i \Delta t, t + \Delta t) = f_i^*(\boldsymbol{x}, t),
\end{equation}
which streams the particle populations to their corresponding neighboring lattice nodes.
\par
The simplest way to account for collisions is to relax the distributions toward their equilibrium $f_i^\text{eq}$, as is done by the Bhatnagar--Gross--Krook (BGK) collision operator \cite{Bhatnagar_Gross_Krook_1954}
\begin{equation}
	\label{eq:bgk}
	\Omega_i(\boldsymbol{x}, t) = - \frac{1}{\tau} (f_i(\boldsymbol{x}, t) - f_i^\text{eq}(\rho_\text{f}, \boldsymbol{u}_\text{f})),
\end{equation}
with the relaxation time $\tau$ that determines the speed of the relaxation and depends on the kinematic viscosity $\nu$ as follows
\begin{equation}
	\label{eq:relaxation}
	\tau = \left( 3 \nu + \frac{1}{2} \right).
\end{equation}
The Maxwell--Boltzmann distribution, quantifying the equilibrium state is given by~\cite{Qian1992}
\begin{equation}
	\label{eq:maxbolt}
	f_i^\text{eq}(\rho_\text{f}, \boldsymbol{u}_\text{f}) = w_i \rho_\text{f} \left( 1 + \frac{\boldsymbol{c}_i \cdot \boldsymbol{u}_\text{f}}{c_s^2} + \frac{(\boldsymbol{c}_i \cdot \boldsymbol{u}_\text{f})^2}{2c_s^4} - \frac{\boldsymbol{u}_\text{f} \cdot \boldsymbol{u}_\text{f}}{2c_s^2} \right).
\end{equation}
The required weights $w_i$ originate from a Gauss-Hermite quadrature rule and are fixed for the chosen velocity set, as is the constant lattice speed of sound $c_s$.
For D$3$Q$19$, the weights read
\begin{equation}
	w_i = 
	\begin{cases}
		1/3, &\text{if } i=0 \\
		1/18, &\text{if } i = 1, \ldots, 6 \\
		1/36, &\text{if } i = 7, \ldots, 18 \\
	\end{cases},
\end{equation}
and the lattice speed of sound is given by $c_s=1/\sqrt{3}$.
\par
The lattice Boltzmann method (LBM) implemented in the open source software OpenLB \cite{olb16,Krause2020} is used exclusively in this work.

\subsection{Homogenized Lattice Boltzmann Method}
\label{sec:methods:psm}

Although the proposed scheme is potentially applicable to a variety of methods such as PSM, in the context of this work, we consider the application to \HLBM{}~\cite{Krause_Klemens_Henn_Trunk_Nirschl_2017,Trunk_Marquardt_Thaeter_Nirschl_Krause_2018,Trunk_Weckerle_Hafen_Thaeter_Nirschl_Krause_2021}. Using the trigonometric level set function \cite{Marquardt_Römer_Nirschl_Krause_2022}
\begin{equation}
	B(\boldsymbol{x}, t) = 
	\begin{cases}
		1, &\text{if } d_s \leq -\varepsilon/2 \\
		\cos^2(\frac{\pi}{2}\left(\frac{d_s}{\varepsilon} + \frac{1}{2}\right), &\text{if } \varepsilon/2 > d_s > -\varepsilon/2 \\
		0, &\text{if } d_s \geq \varepsilon/2 \\
	\end{cases},
\end{equation}
particles are mapped onto the entire computational domain for later coupling between the components, with the size of the smooth boundary $\varepsilon$ and the signed distance to the boundary $d_s$.
Inside the object, the distance is negative, outside it is positive.
In the remainder of the paper, we use $\varepsilon = 1 / 2$, in accordance with Krause et al. \cite{Krause_Klemens_Henn_Trunk_Nirschl_2017}.
The coupling from the particle to the fluid is based on a velocity difference obtained by a convex combination of the fluid and particle velocities
\begin{equation}
	\label{eq:hlbm:du}
	\Delta \boldsymbol{u}_\text{f}(\boldsymbol{x}, t) = B(\boldsymbol{x}, t) \left( \boldsymbol{u}_\text{p}(\boldsymbol{x}, t) - \boldsymbol{u}_\text{f}(\boldsymbol{x}, t) \right),
\end{equation}
where $\boldsymbol{u}_\text{p}(\boldsymbol{x}) = \boldsymbol{u}_\text{p}(\boldsymbol{X}_\text{p}) + \boldsymbol{\omega}_\text{p} \times (\boldsymbol{x} - \boldsymbol{X}_\text{p})$ is the particle's velocity at position $\boldsymbol{x}$, and $\boldsymbol{X}_\text{p}$ is the particle's center of mass.
According to previous studies \cite{Trunk_Weckerle_Hafen_Thaeter_Nirschl_Krause_2021}, the best results are obtained by using an adapted exact difference method (EDM)~\cite{kupershtokh2009EquationsStateLattice}.
This method introduces the following source term in \cref{eq:collide}
\begin{equation}
	\label{eq:kupershtokh}
	S_i(\boldsymbol{x},t) = f_i^\text{eq}(\rho_\text{f}, \boldsymbol{u}_\text{f}+\Delta\boldsymbol{u}_\text{f}) - f_i^\text{eq}(\rho_\text{f}, \boldsymbol{u}_\text{f}).
\end{equation}
To couple from the fluid to the particle, we use the momentum exchange algorithm (MEA) by Wen et al. \cite{wen2014GalileanInvariantFluid} to calculate the hydrodynamic force
\begin{align}
	\label{eq:hlbm:mea}
	\boldsymbol{F}_{\text{h}}(\boldsymbol{x}, t) = 
	\sum_i (\boldsymbol{c}_i - \boldsymbol{u}_\text{p}(\boldsymbol{x}, t)) f_i(\boldsymbol{x} + \boldsymbol{c}_i \Delta t, t) + (\boldsymbol{c}_i + \boldsymbol{u}_\text{p}(\boldsymbol{x}, t)) f_{\bar{i}}(\boldsymbol{x}, t).
\end{align}
Above, the index ${\bar{i}}$ refers to particle populations with the corresponding velocity $\boldsymbol{c}_{\bar{i}} = - \boldsymbol{c}_i$, i.e. the population $f_{\bar{i}}$ points in the opposite direction of $f_{i}$.
\par
The sum of all hydrodynamic forces of cells whose center is inside the particle is now the total hydrodynamic force acting on the particle.
These cells are denoted by $\boldsymbol{x}_\text{b}$.
Thus the total force is~\cite{Trunk_Weckerle_Hafen_Thaeter_Nirschl_Krause_2021}
\begin{align}
	\label{eq:hlbm:hydrodynamicforce}
	\boldsymbol{F}_{\text{p}}(t) = 
	\sum_{\boldsymbol{x}_\text{b}} \boldsymbol{F}_{\text{h}}(\boldsymbol{x}_\text{b}, t),
\end{align}
and the torque is given by
\begin{align}
	\label{eq:hlbm:hydrodynamictorque}
	\boldsymbol{T}_{\text{p}}(t) = 
	\sum_{\boldsymbol{x}_\text{b}} (\boldsymbol{x}_\text{b} - \boldsymbol{X}_\text{p}) \times  \boldsymbol{F}_{\text{h}}(\boldsymbol{x}_\text{b}, t).
\end{align}
Note that while an applicable contact model exists \cite{Marquardt_Römer_Nirschl_Krause_2022}, we deliberately refrain from using it in the context of this work in order to focus exclusively on studying the performance improvement of the decomposition scheme.

\section{Particle Decomposition Scheme}
\label{sec:decomposition}

\subsection{Background}
\label{sec:decomposition:background}

In the context of the LBM, parallelization is typically implemented using a block-based approach \cite{Krueger2016}.
This involves dividing the computational domain into multiple blocks, with each block being assigned to a specific process unit.
In this parallel scheme, a process unit can handle multiple blocks or a single block, depending on the workload distribution.
\par
In the LBM framework, data is stored locally for each block, allowing for efficient computations within each process.
However, as the boundaries of the blocks interact, it becomes necessary to communicate data across these boundaries with directly adjacent blocks.
Implementations of PSMs follow this approach.
The calculation of the surface forces is done locally using the fluid information available within each block.
However, once calculated, these surface forces are communicated to all other processes involved in the simulation for summation.
This allows synchronized solving of the equations of motion on each processor.

\subsection{Improvements}
\label{sec:decomposition:additions}

This section introduces three new steps necessary for particle decomposition, which allow for more efficient communication than the communication between all processes involved as described above.
Importantly, these steps do not change the underlying methodology.
Rather, they balance the workload over multiple processes and ensure data consistency without affecting the accuracy or stability of the method.
They are also applicable to any shape, since the only geometric parameter they depend on is the circumferential radius.
The steps include the communication of surface forces and torques, the assignment of particles, and the communication of particle data.
The following sections provide basic definitions and a detailed explanation of each step and its relevance.

\subsubsection{Definitions}
\label{sec:decomposition:additions:definitions}

To enhance the comprehensibility, we present several definitions that we use in the following sections:
\paragraph{Responsibility} 
When a block or its corresponding process unit assumes responsibility for a particle, it means that both the solving of the equations of motion and the subsequent reassignment of the particle are performed within that specific process.
\paragraph{Neighborhood} 
The neighborhood refers to the collection of blocks located within the maximum circumferential radius, which is the largest radius of all particles in the simulation, from the block of interest.
Neighboring processes correspond to the processes responsible for handling the blocks within this defined block neighborhood.
\paragraph{Extended neighborhood}
The extended neighborhood in this context includes neighboring blocks and their neighboring blocks, creating an extended spatial region that includes not only immediate neighbors but also the secondary level of neighboring blocks.

\subsubsection{Communication of Local Fluid to Particle Coupling Results}
\label{sec:decomposition:additions:communication_forces}
\begin{figure}
	\centering
	\includegraphics[height=5.5cm, width=5.5cm]{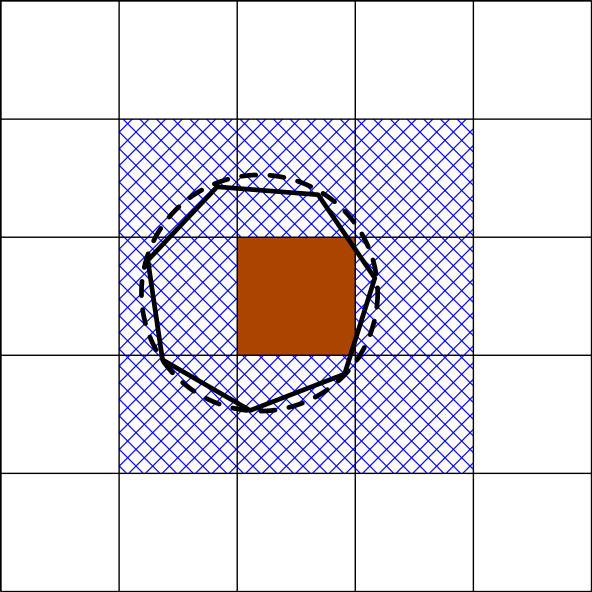}
	\caption{Representation of a heptagonal particle (solid black) and its circumferential radius (dashed black) in a multi-block domain consisting of the responsible center block (brown) and neighboring blocks (blue cross-hatching).
		\label{fig:particle-decomposition}}
\end{figure}

We use a communication-ideal strategy, i.e., particle data is stored and used for coupling on the process units that also know the associated fluid data, due to the advantages outlined by Henn et al. \cite{Henn_Thäter_Dörfler_Nirschl_Krause_2016}. 
Therefore, effective communication of surface forces and torques between process units is required to ensure accurate simulations.
Due to the decomposition of the fluid domains across processors, the responsible processor may lack essential fluid data for the surface parts of a particular particle, as illustrated in \Cref{fig:particle-decomposition}.
In such scenarios, portions of the force and torque are missing and are provided by other processes that have the relevant data, analogously to the conventional implementation.
\par
The new and improved data communication process involves two types of processors: senders and receivers.
The sender processor is the one that holds part of the surface of a particle, but is not responsible for it.
On the other hand, the receiver processor is responsible for the particle in question.
In \Cref{fig:particle-decomposition}, the former is highlighted with a blue cross-hatch and the latter with a brown fill.
\par
The communicated data includes the particle ID, which uniquely identifies the particle within the simulation.
In addition, the sender transmits the partial surface force, which represents the force acting on the surface part of the particle held by the sender.
The torque resulting from the partial surface force is also included.

\subsubsection{Particle Assignment}
\label{sec:decomposition:additions:particle_assignment}

The assignment of particles to processors is done at the block level, similar to the fluid domain decomposition.
This means that a single process may handle multiple particle blocks, depending on the computational domain and decomposition scheme.
After solving the equations of motion, the positions of the particles may have changed. Consequently, the assignment of particles to processors must be reevaluated.
\par
The responsibility for a particle goes to the block on which the center of mass of the particle is located.
Additionally, all neighboring blocks are assumed to touch the surface of the particle.
These neighboring blocks calculate the partial surface forces and torques, but are not individually responsible for the particle itself.
On the other hand, blocks that are not in the neighborhood are not assigned to the particle of interest.
The responsible block described above is shown in \Cref{fig:particle-decomposition} with a brown fill, the neighboring blocks with a blue crosshatch, and all other blocks without fill.
This selectivity ensures an efficient use of computational resources by involving only the necessary blocks in particle interaction calculations and minimizing the following communication overhead by excluding as many process units as possible.
\par
Note that using the maximum circumferential radius to determine the neighborhood for particle assignment avoids missing overlaps and thus ensures data consistency in the communication steps, no matter how complex and different the shapes in the particle system are.

\subsubsection{Communication of Particle Data}
\label{sec:decomposition:additions:communication_particle_data}

\begin{figure}
	\centering
	\includegraphics[height=6.5cm, width=6.5cm]{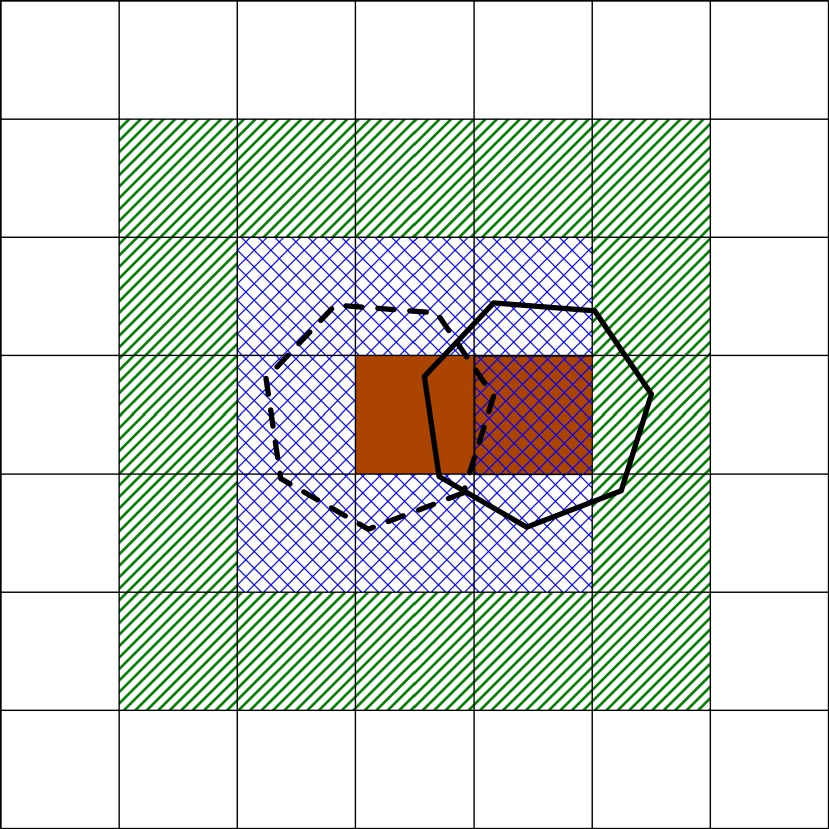}
	\caption{Illustration of the position of a heptagonal structure at the current (dashed black) and next (solid black) time steps within a multi-block domain. The brown middle block represents the currently responsible block, while the adjacent blocks with blue cross-hatching indicate the first-level neighborhood. The highlighted block with a combination of brown fill and blue cross-hatching shows the responsible block at the next time step. The green hatching represents the secondary neighborhood.
		\label{fig:particle-decomposition-relocation}}
\end{figure}

This communication step, which occurs after the particle reassignment, is essential to maintain data consistency and involves more blocks, as shown in \Cref{fig:particle-decomposition-relocation}.
The figure contains a heptagonal particle at the current position and at the next time step, indicated by the dashed and solid lines, respectively.
There are also several types of blocks: the responsible block at the current time step (brown), the responsible block at the next time step (brown with blue cross-hatch), the current neighborhood (blue cross-hatch), and the additional blocks in the current extended neighborhood (green hatching).

The responsible block initiates a request to its entire extended neighborhood to ensure that no surface forces are overlooked due to potential changes in particle responsibility, because, as shown in \Cref{fig:particle-decomposition-relocation}, parts of the particle surface may intersect with the extended neighborhood at the next time step.
While the neighborhood of the block where the particle's center of mass is located after reassignment receives the actual particle data, all other blocks receive an empty request, streamlining the communication process.
\par
The data sent during the communication step includes all relevant particle information.
This typically consists of the particle's unique ID, position, angle, velocity, angular velocity, an ID of the responsible block or process unit, a surface ID, and any other particle properties needed for the specific simulation.
\par
If a block previously held a valid particle but receives no updates in this step, the particle is invalidated and the corresponding memory is released at a predefined interval.

\subsubsection{Time Step Algorithm}
\label{sec:decomposition:additions:algorithm}

To provide a clear and organized presentation, Algorithm~\ref{alg:lbm_psm_with_particle_decomposition} presents the basic time step algorithm of the LBM with PSM and the proposed particle decomposition as it is implemented in the open source software OpenLB \cite{Krause2020}.
The algorithm follows a specific sequence that is consistent with the previously discussed methods and ensures the necessary order of execution.

\begin{algorithm}[H]
	\label{alg:lbm_psm_with_particle_decomposition}
	\caption{\protect Basic LBM time step algorithm using PSM with the particle decomposition scheme}
	\For{all time steps}{
		Couple fluid to particles\Comment*[r]{Using the MEA}
		Communicate surface forces and torques\Comment*[r]{See Section \ref{sec:decomposition:additions:communication_forces}}
		Apply external forces\Comment*[r]{Such as gravity}
		Solve equations of motion\;
		Evaluate particle assignment\Comment*[r]{See Section \ref{sec:decomposition:additions:particle_assignment}}
		Communicate data and assignment\Comment*[r]{See Section \ref{sec:decomposition:additions:communication_particle_data}}
		Couple particles to fluid\Comment*[r]{Using the EDM}
		Perform collision and streaming\;
		Increase time step\;
	}
\end{algorithm}

\section{Application to Hindered Settling}
\label{sec:application}
In the following, we evaluate the application of the proposed scheme to hindered settling \cite{Steinour_1944}.
To ensure the correctness of the results, we validate the average settling velocity, for which many correlations are known.
Most correlations, like the one proposed by Richardson and Zaki~\cite{Richardson_Zaki_1954}, use a power-law model such as
\begin{align}
	\label{eq:avg_vel_power_law}
	\bar{u}_\text{p} = u^*(1 - \phi_\text{p})^n,
\end{align}
with the predicted average settling velocity $\bar{u}_\text{p}$, a reference terminal velocity of a single particle in the considered domain $u^*$, the particle volume fraction $\phi_\text{p}$ and an expansion index
\begin{align}
	\label{eq:n_riza}
	n = 
	\begin{cases}
		4.65, &\text{if } \Rey < 0.2 \\
		4.35 \Rey^{-0.03}, &\text{if } 0.2 \leq \Rey < 1 \\
		4.45 \Rey^{-0.1}, &\text{if } 1 \leq \Rey < 500 \\
		2.39, &\text{if } 500 \leq \Rey
	\end{cases}.
\end{align}
Here, $n$ depends on the Reynolds number $\Rey = u^* D_\text{s} / \nu$.
Garside and Al-Dibouni~\cite{Garside_Al-Dibouni_1977} suggest
\begin{align}
	\label{eq:n_gad}
	n = \frac{5.1 + 0.27 \Rey^{0.9}}{1 + 0.1 \Rey^{0.9}},
\end{align}
which has shown a superior accuracy \cite{Yin_Koch_2007}.
Barnea and Mizrahi~\cite{Barnea_Mizrahi_1973} propose a different model, in which the average settling velocity is independent of $\Rey$ and is given by
\begin{align}
	\label{eq:avg_vel_bm}
	\bar{u}_\text{p} = u^* \frac{1 - \phi_\text{p}}{(1 + \phi_\text{p}^{1/3}) \exp{\frac{5 \phi_\text{p}}{3 (1 - \phi_\text{p})}}}.
\end{align}
For validation purposes, all three correlations are used below.
\par
In the creeping flow regime, the reference velocity reads
\begin{align}
	u^*_\text{St} = \frac{g D_\text{s}^2}{18 \nu} \left(\frac{\rho_\text{p} - \rho_\text{f}}{\rho_\text{f}}\right),
\end{align}
where $g = 9.80665$~\si{m.s^{-2}} is the standard gravity and $D_\text{s}$ is the diameter of the considered spherical particle.
Furthermore, $\rho_\text{f}$ and $\rho_\text{p}$ are the fluid and particle densities.
However, in the following studies we only consider higher Reynolds numbers, hence we calculate the reference velocity using
\begin{align}
	u^* = \sqrt{\frac{4 g D_\text{s}}{3 C_d} \left( \frac{\rho_\text{p} - \rho_\text{f}}{\rho_\text{f}} \right)},
\end{align}
with the drag coefficient $C_d$ that we compute using the approximation by Schiller and Neumann~\cite{schiller1933uber}
\begin{align}
	C_d = \frac{24}{\Rey} \left( 1 + 0.15 \Rey^{0.687} \right),
\end{align}
which is valid for $\Rey < 800$.
\par
As is common in studies of hindered settling~\cite{Yao_Criddle_Fringer_2021, Yin_Koch_2007}, we use two dimensionless parameters to describe the setup below.
The first one is the particle to fluid density ratio defined as $\rho_\text{p}/\rho_\text{f}$ and the second is the Archimedes number
\begin{align}
	Ar = \frac{g D_\text{s}^3 \frac{\rho_\text{p} - \rho_\text{f}}{\rho_\text{f}}}{\nu^2}.
\end{align}

\subsection{Simulation Setup}
\label{sec:application:setup}
\begin{figure}
	\centering
	\includegraphics{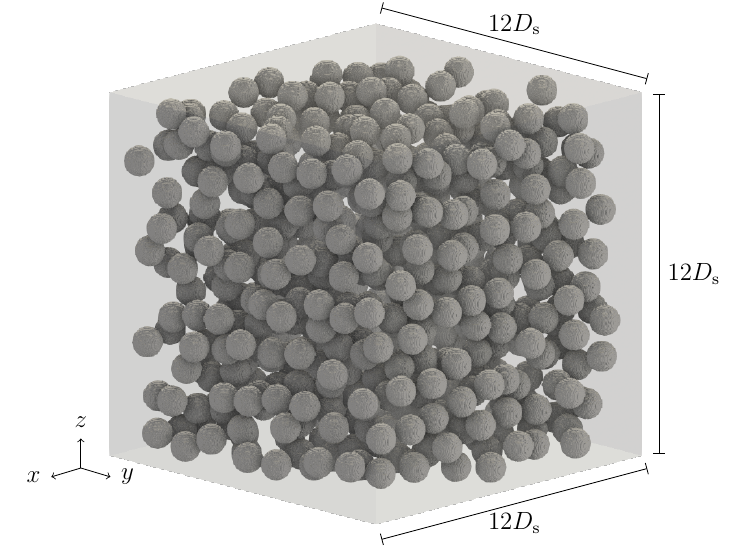}
	
	\caption{Simulation domain with the spherical particles on their initial position using a resolution of $N=27$ cells per sphere diameter and particle volume fraction of about $0.2$.
		\label{fig:hindered_settling_setup_example}}
\end{figure}

To numerically study the above, we employ spherical particles with a diameter $D_\text{s} = 2$~\si{\milli\meter} at random positions in a cubic domain with an edge length of $12D_\text{s}$ and periodic boundaries on each side, which is filled with a fluid with a density $\rho_\text{f} = 1000$~\si{\kilogram\per\cubic\meter}.
The setup is exemplified in \Cref{fig:hindered_settling_setup_example} with a particle volume fraction of about $0.2$.
The size of the domain was chosen because previous studies have shown that sufficiently accurate results can be obtained from a size of $10D_\text{s}$~\cite{Yao_Criddle_Fringer_2021, Yin_Koch_2007}.
However, in the context of this work, we intend to generate larger particle numbers for testing purposes, and therefore expand the domain, but only slightly, so that the simulation remains feasible for high resolutions.

The initially resting particles are accelerated in the $z$-direction by the force $F_g = (\rho_\text{f}~-~\rho_\text{p}) V_\text{s} g$.
Here, $V_\text{s}~= \pi D_\text{s}^3 / 6$ is the sphere's volume.
As in previous numerical investigations of hindered settling \cite{Willen_Prosperetti_2019,Yao_Criddle_Fringer_2021,Yin_Koch_2007}, we apply a pressure gradient in the opposite direction of particle motion to obtain a net volume flow rate of zero.
We calculate it using the Ergun equation~\cite{Ergun_1952} with a superficial velocity equal to the current average settling velocity $\bar{u}_\text{p}$.
\par
In our studies, the Archimedes number is fixed at $Ar = 1500$ and the density ratio at $\rho_\text{p}/\rho_\text{f} = 1.3$.
From these two quantities, we derive the particle density as well as the fluid viscosity.
To explore the effects of particle concentration, we vary the particle volume fraction in the range of 0.05 to 0.3.
\par
Furthermore, we choose a constant lattice relaxation time of $0.55$ for all simulations, while resolving the sphere's diameter with $N$ cells, which differs in the following simulations.
We simulate a total time of $300 t^*$, with $t^* = D_\text{s} / u^*_\text{St}$.
Due to the initial random packing, the startup time is significantly reduced, nonetheless, we start averaging the velocities to obtain $\bar{u}_\text{p}$ after a time of $30 t^*$, leaving a period of $270 t^*$ that ensures statistically averaged results.

\subsection{Grid Independence Study}
\label{sec:application:grid_independence}

\begin{figure}
	\centering
	\includegraphics[height=6.5cm, width=10.5cm]{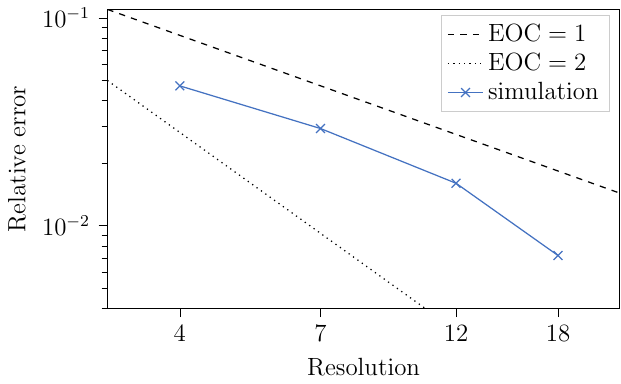}
	\caption{Relative error in $L^2$ norm versus the resolution of the sphere's diameter $N$. The reference uses the resolution $N=27$.
		\label{fig:grid_independence_avg_velocity_error}}
\end{figure}

In this section, we delve into a grid independence study using a particle volume fraction of $\phi_\text{p} = 0.1$ by systematically exploring various grid resolutions $N \in \{5, 7, 12, 18\}$.
We compare the outcomes obtained from these resolutions to evaluate their impact on the simulation results, with respect to a baseline resolution of $N=27$.
Previous studies report sufficiently accurate results when resolving a sphere diameter with $N \approx 8$ cells~\cite{Trunk_Weckerle_Hafen_Thaeter_Nirschl_Krause_2021}.
In order to have a very accurate baseline, we therefore use a much finer resolution while still ensuring feasible simulations.
The results are presented in \Cref{fig:grid_independence_avg_velocity_error}, which illustrates the relative error using the $L^2$ norm \cite{Krueger2016} plotted against the used grid resolutions.
Additionally, the figure includes lines representing the experimental orders of convergence (EOC) with values of $1$ and $2$.

The plot reveals that as the grid resolution increases, the error of the simulation approximately follows the line of the EOC of 1, indicating a linear decrease in error with respect to grid refinement.
The error decreases with increasing resolution as the particle shape is represented more accurately.
The observed linear convergence is in agreement with the results of previous studies on single settling spheres~\cite{Trunk_Weckerle_Hafen_Thaeter_Nirschl_Krause_2021} and suggests that increasing the grid resolution leads to a proportional decrease in error.
It is likely due to the staircase approximation of the curved boundary~\cite{Krueger2016,Ginzbourg1996}.
Moreover, it is noteworthy that the error becomes smaller than $2$\% starting from a resolution of $N=12$, ensuring grid-independent results.
The minimum resolution required is slightly higher than the resolutions reported in previous studies~\cite{Trunk_Weckerle_Hafen_Thaeter_Nirschl_Krause_2021}.
This adjustment is due to the presence of multiple particles and their interactions through the fluid.

\subsection{Validation}
\label{sec:application:validation}

\begin{figure}
	\centering
	\includegraphics[height=8.0cm, width=10.0cm]{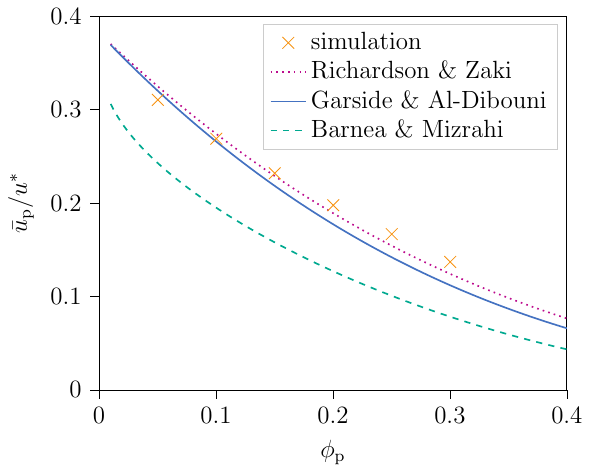}
	\caption{Ratio of average settling velocity to single particle settling velocity $\bar{u}_\text{p} / u^*$ versus the particle volume fraction $\phi_\text{p}$, showing current simulation results and the correlations by Richardson and Zaki~\cite{Richardson_Zaki_1954}, Garside and Al-Dibouni~\cite{Garside_Al-Dibouni_1977}, as well as Barnea and Mizrahi~\cite{Barnea_Mizrahi_1973}.
	\label{fig:hindered_settling_validation}}
\end{figure}

The following validation uses the resolution $N=18$, to assess the accuracy and reliability.
To this end, we plot the average velocity obtained from the simulations over the particle volume fractions in \Cref{fig:hindered_settling_validation}.
For comparison, we additionally add the correlation by Richardson and Zaki~\cite{Richardson_Zaki_1954} as a dotted purple line, Garside and Al-Dibouni~\cite{Garside_Al-Dibouni_1977} as a blue line, and the correlation by Barnea and Mizrahi~\cite{Barnea_Mizrahi_1973} as a dashed green line.
It is evident that as the particle volume fraction increases, the average settling velocity progressively decreases.
In general, the results are well within the range of the correlations introduced above.
In particular, they agree well with the first two correlations.
However, it is noteworthy that at higher particle volume fractions, the disparity between the simulation results and correlations becomes more pronounced.

\HLBM{} is therefore able to capture the settling behavior at low particle volume fractions, suggesting that in these cases \HLBM{} works without an explicit contact model, since the interaction through the fluid is prominent.
The observed discrepancy between the simulation results and the correlations at higher particle volume fractions can be attributed to the absence of an explicit contact model, because with increasing particle volume fraction, the probability of particle-particle interactions also increases.
This in accordance with previous findings~\cite{Trunk_Weckerle_Hafen_Thaeter_Nirschl_Krause_2021}.
In simulations without an explicit contact model, these interactions are not fully accounted for, leading to an error in the predicted settling velocities.
Since the energy dissipation due to interparticle interactions is missing, the observed overestimation of the velocity seems reasonable.
The inclusion of a contact model \cite{Marquardt_Römer_Nirschl_Krause_2022,Karimnejad2022} would allow for a more accurate representation of particle-particle interactions, potentially reducing the difference between simulation results and correlations, especially at higher particle volume fractions.
However, since we are primarily interested in quantifying the performance improvement from decomposition, an elaborate contact model would introduce too much bias by degrading overall performance by introducing more complex computation and communication.
\par
\begin{figure}
	\captionsetup[subfigure]{justification=centering,aboveskip=-1pt,belowskip=-1pt}
	\centering
	\begin{subfigure}[b]{\textwidth}
		\centering
		\includegraphics[height=0.95cm]{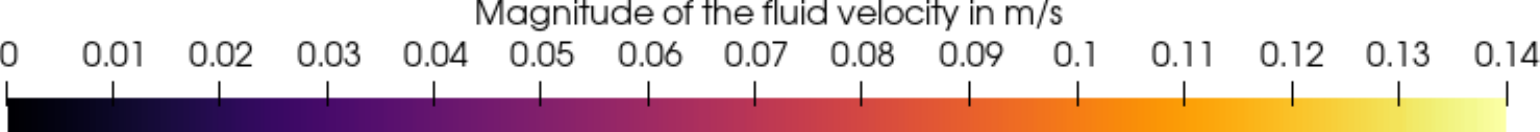}
	\end{subfigure}
	\newline
	\vspace{-0.2cm}
	
	\begin{subfigure}[b]{0.475\textwidth}
		\centering
		\includegraphics[height=4.7cm]{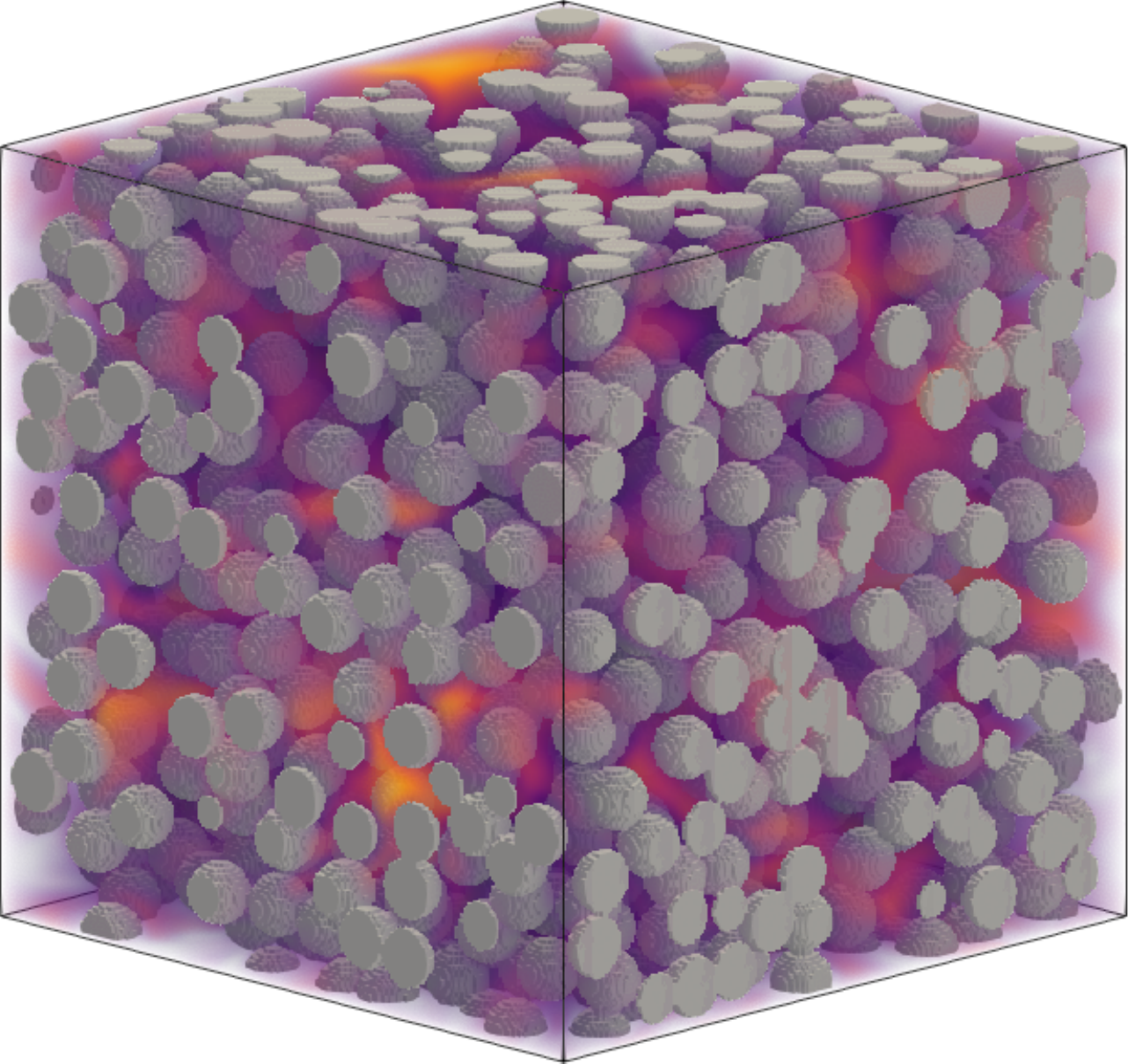}
		
		\caption{$t^* \approx 49.97$}  
		\label{fig:sim-vis-spheres-2}
	\end{subfigure}
	\hfill
	\begin{subfigure}[b]{0.475\textwidth}
		\centering
		\includegraphics[height=4.7cm]{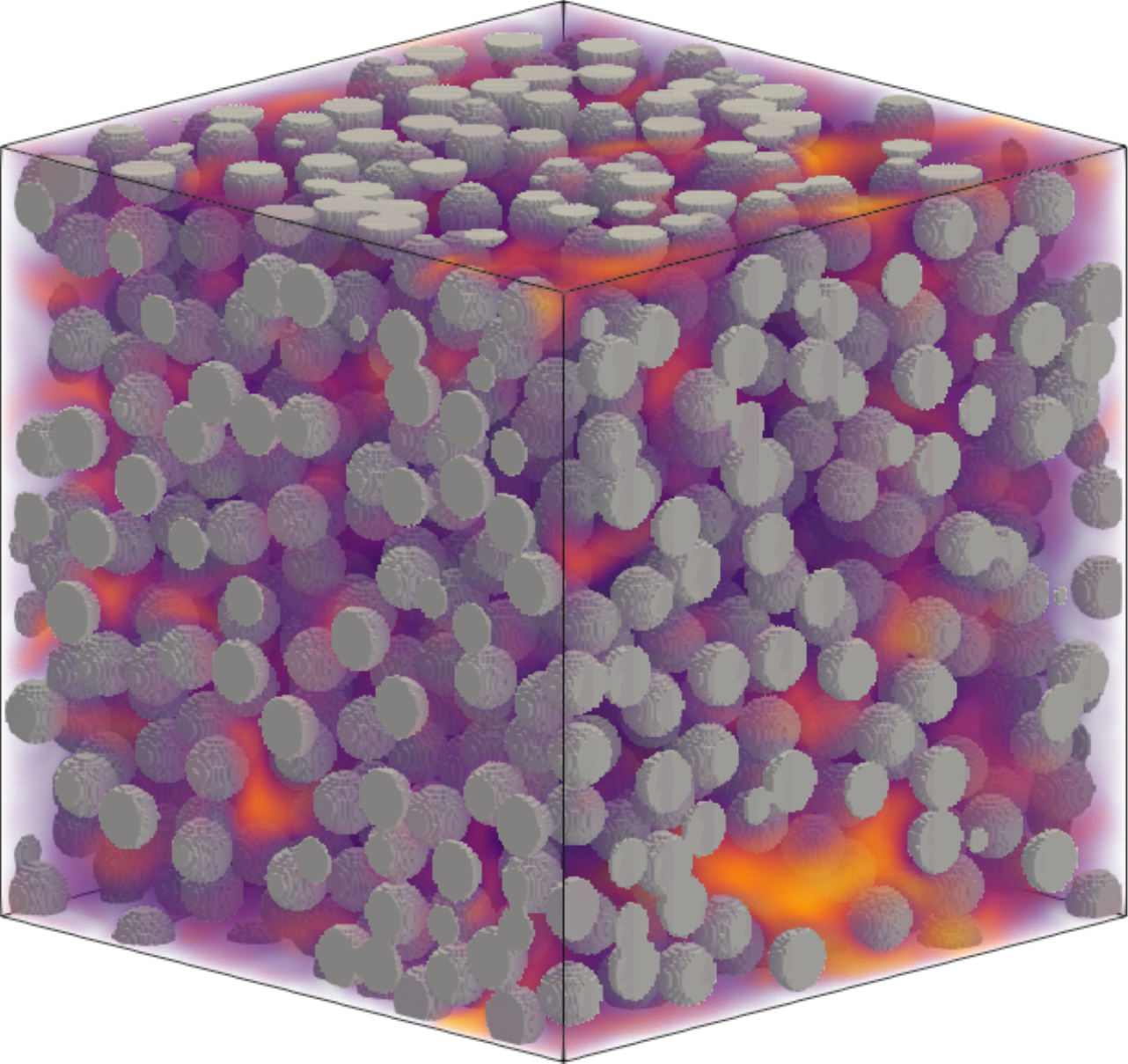}
		
		\caption{$t^* \approx 299.82$}  
		\label{fig:sim-vis-cubes-2}
	\end{subfigure}
	
	\caption{Particles and fluid velocity at different normalized times $t^*$ for $Ar = 1500$ and $\phi_\text{p} \approx 0.3$.
		\label{fig:hindered_settling_validation_visualization_of_simulation}}
\end{figure}

\Cref{fig:hindered_settling_validation_visualization_of_simulation} visualizes the fluid velocity and the particles for $\phi_\text{p} \approx 0.3$ at different normalized times $t^*$.
It can be seen that the particles form clusters and that high fluid velocity channels form.
The observation is quantitatively consistent with the literature~\cite{Yao_Criddle_Fringer_2021}, further confirming the correctness.
However, interparticle contacts are also visible, hinting at the potential need for an explicit contact model at high particle volume fractions.

\subsection{Performance}
\label{sec:application:performance}

In this section, we evaluate the performance of \HLBM{} with the novel particle decomposition scheme and the conventional one.
The computational infrastructure used for these experiments consists of Intel Xeon Platinum 8368 CPUs, with each node equipped with 76 CPU cores.
\par
\begin{figure}
	\captionsetup[subfigure]{justification=centering,aboveskip=-1pt,belowskip=-1pt}
	\centering
	\begin{subfigure}[b]{0.6\textwidth}
		\centering
		\includegraphics[height=1.2cm]{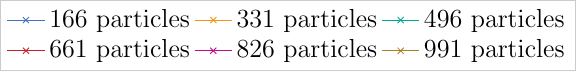}
	\end{subfigure}
	\par
	
	\begin{subfigure}[b]{0.475\textwidth}
		\centering
		\includegraphics[height=4.4cm,width=\textwidth]{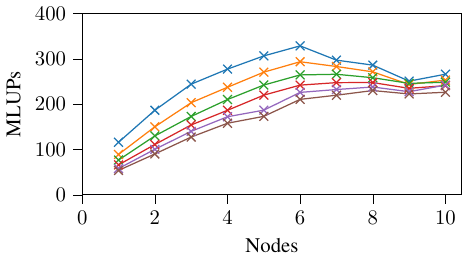}
		\caption{} 
		\label{fig:particle-decomposition-performance-comparison-n=12-with}
	\end{subfigure}
	\hfill
	\begin{subfigure}[b]{0.475\textwidth}
		\centering
		\includegraphics[height=4.4cm,width=\textwidth]{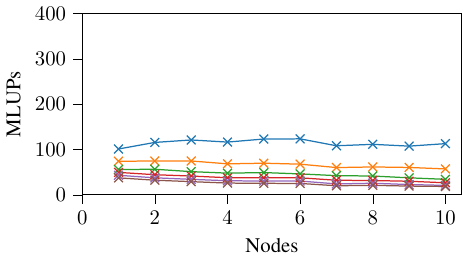}
		\caption{} 
		\label{fig:particle-decomposition-performance-comparison-n=12-without}
	\end{subfigure}
	\par
	\begin{subfigure}[b]{0.475\textwidth}
		\centering
		\includegraphics[height=4.4cm,width=\textwidth]{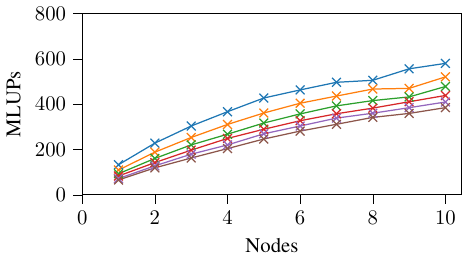}
		\caption{} 
		\label{fig:particle-decomposition-performance-comparison-n=18-with}
	\end{subfigure}
	\hfill
	\begin{subfigure}[b]{0.475\textwidth}
		\centering
		\includegraphics[height=4.4cm,width=\textwidth]{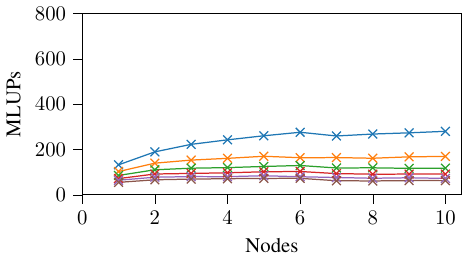}
		\caption{} 
		\label{fig:particle-decomposition-performance-comparison-n=18-without}
	\end{subfigure}
	\par
	\begin{subfigure}[b]{0.475\textwidth}
		\centering
		\includegraphics[height=4.4cm,width=\textwidth]{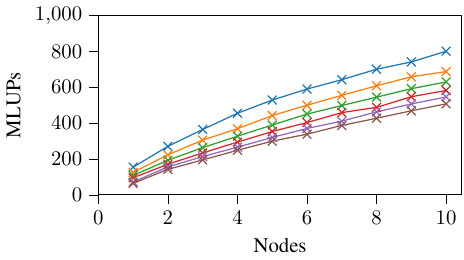}
		\caption{} 
		\label{fig:particle-decomposition-performance-comparison-n=27-with}
	\end{subfigure}
	\hfill
	\begin{subfigure}[b]{0.475\textwidth}
		\centering
		\includegraphics[height=4.4cm,width=\textwidth]{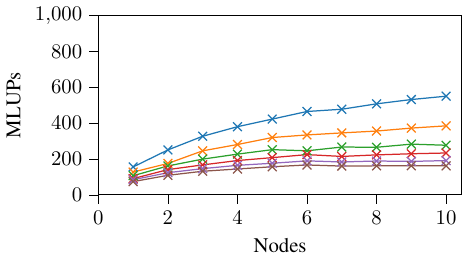}
		\caption{} 
		\label{fig:particle-decomposition-performance-comparison-n=27-without}
	\end{subfigure}
	
	\caption{Comparison of the resulting MLUPs versus the number of employed nodes using the new particle decomposition scheme (left) and the conventional decomposition (right) for the resolutions $N=12$ (\textbf{a} and \textbf{b}), $N=18$ (\textbf{c} and \textbf{d}) and $N=27$ (\textbf{e} and \textbf{f}) and different particle numbers.
		\label{fig:particle-decomposition-performance-comparison}}
\end{figure}

\Cref{fig:particle-decomposition-performance-comparison} presents the results of this analysis.
Here, we plot the million lattice site updates per second  (MLUPs) versus the number of nodes used.
We consider a fixed total problem size with a resolution of $12$ in \Cref{fig:particle-decomposition-performance-comparison-n=12-with,fig:particle-decomposition-performance-comparison-n=12-without}, a resolution of $18$ in \Cref{fig:particle-decomposition-performance-comparison-n=18-with,fig:particle-decomposition-performance-comparison-n=18-without}, and a resolution of $27$ in \Cref{fig:particle-decomposition-performance-comparison-n=27-with,fig:particle-decomposition-performance-comparison-n=27-without}.
In each scenario, the number of particles ranges from 166 to 991. 

The performance plots in \Cref{fig:particle-decomposition-performance-comparison} reveal several important observations:
Firstly, the novel method exhibits in general higher MLUPs compared to the conventional method.
However, for small resolution and low particle number, the MLUPs are of a similar scale, but as particle number and resolution increase, the differences increase.
For example, the simulations with the new particle decomposition show MLUPs more than five times larger than those with the conventional method, when considering $991$ particles.
Additionally, the conventional method demonstrates significant communication limitations for $N=12$, which is immediately apparent by a flattening to a plateau in the plots, and for $N=18$, which become noticeable when the number of nodes reaches six.
\par
In contrast, the new method shows significant communication limitations only for $N=12$ and when the number of nodes reaches seven or more, see \Cref{fig:particle-decomposition-performance-comparison-n=12-with}.
Note that the problem size is smallest in this particular scenario.
Increasing the number of processes beyond the observed inflection point degrades performance because communication becomes the limiting factor.
\par
In a broader context, increasing the number of processes increases the number of costly communications, which explains the flattening of all performance curves as more nodes are used.
At some point, communications become more expensive than computations, leading to the inflection point observed above.
This behavior would occur in all cases, but shows up earlier for small problem sizes because the overall workload is smaller.
\par
The new method demonstrates improved computational efficiency and scalability.
These improvements are due to the more efficient communication strategy of the proposed method.
Unlike the conventional method, which transmits all data to every process, the novel method selectively communicates data within the neighborhood of each block, as explained earlier.
By minimizing unnecessary data transfer and concentrating communication on relevant blocks, the novel method significantly reduces communication overhead and optimizes computational resources.
\par
These results highlight the benefits of the novel method in mitigating communication bottlenecks.
\par
The studies focus on spheres, but a performance gain is expected for any other shape, because in these cases the signed distance function is more complex.
The increased complexity means that the computations performed on each lattice node, see Section~\ref{sec:methods:psm}, are more expensive.
It is therefore even more important to distribute the computational effort across the processes in cases of complex shapes. 
The relative performance gain is expected to be even higher.
\par
The studies furthermore consider a minimum of $166$ particles.
However, the performance improvements are expected to also extend to single particle simulations, due to a reduction in communication overhead, although to a lesser extent.
This reduction is due to the fact that only the processes adjacent to the singular particle are required to communicate, allowing the remaining processes to perform other computations in parallel.

\section{Summary and Conclusions}
\label{sec:summary}

In the present work, we propose and validate a novel and improved particle decomposition scheme for surface resolved particle simulations and apply it to \HLBM{} for demonstration.
For this purpose, we perform extensive studies using the example of hindered settling.
\par
The main objective, performance improvement, is evaluated and confirmed, as the comparison of the conventional and the novel methods in terms of computational efficiency and communication constraints reveals significant advantages offered by the novel method.
The novel method scales considerably better, resulting in higher MLUPs compared to the conventional method.
This improvement is due to the more efficient communication strategy of the novel method, where data is selectively communicated.
\par
The studies also show that \HLBM{} is capable of capturing the settling behavior of a particle swarm at low particle volume fractions.
This suggests that under these conditions, \HLBM{} works effectively without the need for a dedicated contact model.
However, as particle volume fractions increase, the limitations of not having a dedicated contact model become more pronounced.
Therefore, at higher particle volume fractions, there is a clear need to incorporate a dedicated contact model to accurately simulate and account for particle interactions.
\par
These results emphasize the importance of efficient communication strategies in particle simulation methods and highlight the value of the novel method in achieving computational efficiency.
It is now feasible to simulate a larger number of surface resolved particles, resulting in more realistic simulation setups and expanded applicability to real-world problems, such as filters and thickeners.

% Nomenclature
\section*{Nomenclature}
\setlength{\LTleft}{0pt}
\noindent
\textbf{Acronyms}
\begin{longtable}{p{1.1cm}p{9.5cm}}
	BGK & Bhatnagar--Gross--Krook \\
	DEM & discrete element method \\
	EDM & exact difference method \\
	EOC & experimental order of convergence \\
	HLBM & homogenized lattice Boltzmann method \\
	IBM & immersed boundary method \\
	LBM & lattice Boltzmann method \\
	MEA & momentum exchange algorithm \\
	MLUPs & million lattice site updates per second \\
	PSM & partially saturated method \\
\end{longtable}
\noindent
\textbf{Roman Symbols}
\begin{longtable}{p{1.1cm}p{9.5cm}}
	$B$ & weighting factor \\
	$\boldsymbol{c}$ & discrete velocity \\
	$c_s$ & lattice speed of sound \\
	$D$ & diameter \\
	$d_\text{s}$ & signed distance \\
	$\boldsymbol{F}$ & total force \\
	$F_g$ & combination of weight and buoyancy \\
	$f$ & particle population \\
	$f^*$ & post-collision particle population \\
	$g$ & standard gravity \\
	$\boldsymbol{I}$ & moment of inertia \\
	$m$ & mass \\
	$N$ & resolution \\
	$n$ & expansion index \\
	$\Rey$ & Reynolds number \\
	$S$ & source term \\
	$\boldsymbol{T}$ & total torque \\
	$t$ & time \\
	$\boldsymbol{u}$ & velocity \\
	$u^*$ & reference settling velocity of a single sphere \\
	$V$ & volume \\
	$w$ & weight for the equilibrium distribution calculation \\
	$\boldsymbol{X}$ & center of mass \\
	$\boldsymbol{x}$ & position \\
\end{longtable}
\noindent
\textbf{Greek Symbols}
\begin{longtable}{p{1.1cm}p{9.5cm}}
	$\Delta t$ & time step size \\
	$\Delta x$ & grid spacing \\
	$\varepsilon$ & size of the smooth boundary \\
	$\nu$ & kinematic viscosity \\
	$\rho$ & density \\
	$\tau$ & relaxation time \\
	$\phi$ & volume fraction \\
	$\Omega$ & collision operator \\
	$\boldsymbol{\omega}$ & angular velocity \\
\end{longtable}
\noindent
\textbf{Subscripts}
\begin{longtable}{p{1.1cm}p{9.5cm}}
	$\text{b}$ & refers to positions inside a particle's boundary \\
	$\text{f}$ & refers to the fluid \\
	$\text{h}$ & refers to the hydrodynamic force \\
	$i$ & refers to the corresponding discrete velocity \\
	$\text{p}$ & refers to the particle's center of mass \\
	$\text{s}$ & refers to a sphere \\
	$\text{St}$ & refers to the Stokes/creeping flow regime \\
\end{longtable}

\textbf{\emph{Acknowledgements:}}
Funded by the Deutsche Forschungsgemeinschaft (DFG, German Research Foundation) – 422374351. This work was performed on the HoreKa supercomputer funded by the Ministry of Science, Research and the Arts Baden-Württemberg and by the Federal Ministry of Education and Research. 

\textbf{\emph{Author contribution statement:}}
\textbf{J.\ E.\ Marquardt}: Conceptualization, Methodology, Software, Validation, Formal analysis, Investigation, Data curation, Writing - Original Draft, Writing - Review \& Editing, Visualization;
\textbf{N.\ Hafen}: Conceptualization, Methodology, Software, Writing - Review \& Editing, Project administration; 
\textbf{M.\ J.\ Krause}: Software, Resources, Supervision, Project administration, Funding acquisition. 

 %\bibliographystyle{unsrturl} 
 %\bibliography{refs}

\begin{thebibliography}{10}
	
	\bibitem{Steinour_1944}
	Harold~H. Steinour.
	\newblock Rate of sedimentation: Nonflocculated suspensions of uniform spheres.
	\newblock {\em Industrial \& Engineering Chemistry}, 36(7):618–624, Jul 1944.
	\newblock \href {https://doi.org/10.1021/ie50415a005}
	{\path{doi:10.1021/ie50415a005}}.
	
	\bibitem{Richardson_Zaki_1954}
	J.~F. Richardson and W.~N. Zaki.
	\newblock Sedimentation and fluidisation: {Part I}.
	\newblock {\em Chemical Engineering Research and Design}, 75:S82–S100, 1954.
	\newblock \href {https://doi.org/10.1016/S0263-8762(97)80006-8}
	{\path{doi:10.1016/S0263-8762(97)80006-8}}.
	
	\bibitem{Barnea_Mizrahi_1973}
	E.~Barnea and J.~Mizrahi.
	\newblock A generalized approach to the fluid dynamics of particulate systems.
	\newblock {\em The Chemical Engineering Journal}, 5(2):171–189, Jan 1973.
	\newblock \href {https://doi.org/10.1016/0300-9467(73)80008-5}
	{\path{doi:10.1016/0300-9467(73)80008-5}}.
	
	\bibitem{Garside_Al-Dibouni_1977}
	John Garside and Maan~R Al-Dibouni.
	\newblock Velocity-voidage relationships for fluidization and sedimentation in
	solid-liquid systems.
	\newblock {\em Industrial \& engineering chemistry process design and
		development}, 16(2):206–214, 1977.
	\newblock Citation Key: garside1977velocity.
	
	\bibitem{Di_Felice_1995}
	Renzo Di~Felice.
	\newblock Hydrodynamics of liquid fluidisation.
	\newblock {\em Chemical Engineering Science}, 50(8):1213–1245, Apr 1995.
	\newblock \href {https://doi.org/10.1016/0009-2509(95)98838-6}
	{\path{doi:10.1016/0009-2509(95)98838-6}}.
	
	\bibitem{Di_Felice_1999}
	R~Di~Felice.
	\newblock The sedimentation velocity of dilute suspensions of nearly monosized
	spheres.
	\newblock {\em International Journal of Multiphase Flow}, 25(4):559–574, Jun
	1999.
	\newblock \href {https://doi.org/10.1016/S0301-9322(98)00084-6}
	{\path{doi:10.1016/S0301-9322(98)00084-6}}.
	
	\bibitem{Hamid_Molina_Yamamoto_2014}
	Adnan Hamid, John~J. Molina, and Ryoichi Yamamoto.
	\newblock Direct numerical simulations of sedimenting spherical particles at
	non-zero reynolds number.
	\newblock {\em RSC Advances}, 4(96):53681–53693, 2014.
	\newblock \href {https://doi.org/10.1039/C4RA11025K}
	{\path{doi:10.1039/C4RA11025K}}.
	
	\bibitem{Zaidi_Tsuji_Tanaka_2015}
	Ali~Abbas Zaidi, Takuya Tsuji, and Toshitsugu Tanaka.
	\newblock Hindered settling velocity \& structure formation during particle
	settling by direct numerical simulation.
	\newblock {\em Procedia Engineering}, 102:1656–1666, Jan 2015.
	\newblock \href {https://doi.org/10.1016/j.proeng.2015.01.302}
	{\path{doi:10.1016/j.proeng.2015.01.302}}.
	
	\bibitem{Willen_Prosperetti_2019}
	Daniel~P. Willen and Andrea Prosperetti.
	\newblock Resolved simulations of sedimenting suspensions of spheres.
	\newblock {\em Physical Review Fluids}, 4(1):014304, Jan 2019.
	\newblock \href {https://doi.org/10.1103/PhysRevFluids.4.014304}
	{\path{doi:10.1103/PhysRevFluids.4.014304}}.
	
	\bibitem{Yao_Criddle_Fringer_2021}
	Yinuo Yao, Craig~S. Criddle, and Oliver~B. Fringer.
	\newblock The effects of particle clustering on hindered settling in
	high-concentration particle suspensions.
	\newblock {\em Journal of Fluid Mechanics}, 920:A40, Aug 2021.
	\newblock \href {https://doi.org/10.1017/jfm.2021.470}
	{\path{doi:10.1017/jfm.2021.470}}.
	
	\bibitem{Chong_Ratkowsky_Epstein_1979}
	Y.S. Chong, D.A. Ratkowsky, and N.~Epstein.
	\newblock Effect of particle shape on hindered settling in creeping flow.
	\newblock {\em Powder Technology}, 23(1):55–66, May 1979.
	\newblock \href {https://doi.org/10.1016/0032-5910(79)85025-1}
	{\path{doi:10.1016/0032-5910(79)85025-1}}.
	
	\bibitem{Jirout_Jiroutova_2022}
	Tomáš Jirout and Dita Jiroutová.
	\newblock Hindered settling of fiber particles in viscous fluids.
	\newblock {\em Processes}, 10(99):1701, Sep 2022.
	\newblock \href {https://doi.org/10.3390/pr10091701}
	{\path{doi:10.3390/pr10091701}}.
	
	\bibitem{Tomkins_Baldock_Nielsen_2005}
	Matt~R. Tomkins, Tom~E. Baldock, and Peter Nielsen.
	\newblock Hindered settling of sand grains.
	\newblock {\em Sedimentology}, 52(6):1425–1432, 2005.
	\newblock \href {https://doi.org/10.1111/j.1365-3091.2005.00750.x}
	{\path{doi:10.1111/j.1365-3091.2005.00750.x}}.
	
	\bibitem{Zhu_Zhou_Yang_Yu_2008}
	H.~P. Zhu, Z.~Y. Zhou, R.~Y. Yang, and A.~B. Yu.
	\newblock Discrete particle simulation of particulate systems: A review of
	major applications and findings.
	\newblock {\em Chemical Engineering Science}, 63(23):5728–5770, Dec 2008.
	\newblock \href {https://doi.org/10.1016/j.ces.2008.08.006}
	{\path{doi:10.1016/j.ces.2008.08.006}}.
	
	\bibitem{Nolan_Kavanagh_1995}
	G.~T. Nolan and P.~E. Kavanagh.
	\newblock Random packing of nonspherical particles.
	\newblock {\em Powder Technology}, 84(3):199–205, Sep 1995.
	\newblock \href {https://doi.org/10.1016/0032-5910(95)98237-S}
	{\path{doi:10.1016/0032-5910(95)98237-S}}.
	
	\bibitem{Rakotonirina_Delenne_Radjai_Wachs_2019}
	Andriarimina~Daniel Rakotonirina, Jean-Yves Delenne, Farhang Radjai, and
	Anthony Wachs.
	\newblock Grains3D, a flexible {DEM} approach for particles of arbitrary convex
	shape—part {III}: extension to non-convex particles modelled as glued
	convex particles.
	\newblock {\em Computational Particle Mechanics}, 6(1):55–84, 2019.
	\newblock \href {https://doi.org/10.1007/s40571-018-0198-3}
	{\path{doi:10.1007/s40571-018-0198-3}}.
	
	\bibitem{Qiu_Wu_2014}
	Liu-Chao Qiu and Chuan-Yu Wu.
	\newblock A hybrid {DEM}/{CFD} approach for solid-liquid flows.
	\newblock {\em Journal of Hydrodynamics}, 26:19–25, Feb 2014.
	\newblock \href {https://doi.org/10.1016/S1001-6058(14)60003-2}
	{\path{doi:10.1016/S1001-6058(14)60003-2}}.
	
	\bibitem{Sun_Xiao_2016}
	Rui Sun and Heng Xiao.
	\newblock Sedifoam: A general-purpose, open-source {CFD}–{DEM} solver for
	particle-laden flow with emphasis on sediment transport.
	\newblock {\em Computers \& Geosciences}, 89:207–219, Apr 2016.
	\newblock \href {https://doi.org/10.1016/j.cageo.2016.01.011}
	{\path{doi:10.1016/j.cageo.2016.01.011}}.
	
	\bibitem{weers2022DevelopmentModelSeparation}
	Martin Weers, Leonard Hansen, Daniel Schulz, Bernd Benker, Annett Wollmann,
	Carsten Kykal, Harald Kruggel-Emden, and Alfred~P. Weber.
	\newblock Development of a {Model} for the {Separation} {Characteristics} of a
	{Deflector} {Wheel} {Classifier} {Including} {Particle} {Collision} and
	{Rebound} {Behavior}.
	\newblock {\em Minerals}, 12(4):480, April 2022.
	\newblock \href {https://doi.org/10.3390/min12040480}
	{\path{doi:10.3390/min12040480}}.
	
	\bibitem{andersson2011computational}
	Bengt Andersson, Ronnie Andersson, Love H{\aa}kansson, Mikael Mortensen, Rahman
	Sudiyo, and Berend Van~Wachem.
	\newblock {\em Computational fluid dynamics for engineers}.
	\newblock Cambridge university press, 2011.
	\newblock \href {https://doi.org/10.1017/CBO9781139093590}
	{\path{doi:10.1017/CBO9781139093590}}.
	
	\bibitem{Uhlmann_2005}
	Markus Uhlmann.
	\newblock An immersed boundary method with direct forcing for the simulation of
	particulate flows.
	\newblock {\em Journal of Computational Physics}, 209(2):448–476, Nov 2005.
	\newblock \href {https://doi.org/10.1016/j.jcp.2005.03.017}
	{\path{doi:10.1016/j.jcp.2005.03.017}}.
	
	\bibitem{AmiriDelouei2022}
	Amin Amiri~Delouei, Sajjad Karimnejad, and Fuli He.
	\newblock Direct numerical simulation of pulsating flow effect on the
	distribution of non-circular particles with increased levels of complexity:
	{IB}-{LBM}.
	\newblock {\em Computers \& Mathematics with Applications}, 121:115--130,
	September 2022.
	\newblock \href {https://doi.org/10.1016/j.camwa.2022.07.005}
	{\path{doi:10.1016/j.camwa.2022.07.005}}.
	
	\bibitem{Karimnejad2018}
	S.~Karimnejad, A.~Amiri~Delouei, M.~Nazari, M.M. Shahmardan, and A.A. Mohamad.
	\newblock Sedimentation of elliptical particles using immersed boundary –
	lattice boltzmann method: A complementary repulsive force model.
	\newblock {\em Journal of Molecular Liquids}, 262:180--193, July 2018.
	\newblock \href {https://doi.org/10.1016/j.molliq.2018.04.075}
	{\path{doi:10.1016/j.molliq.2018.04.075}}.
	
	\bibitem{Hu2015}
	Yang Hu, Decai Li, Shi Shu, and Xiaodong Niu.
	\newblock Modified momentum exchange method for fluid-particle interactions in
	the lattice boltzmann method.
	\newblock {\em Physical Review E}, 91(3):033301, March 2015.
	\newblock \href {https://doi.org/10.1103/physreve.91.033301}
	{\path{doi:10.1103/physreve.91.033301}}.
	
	\bibitem{Noble_Torczynski_1998}
	D.~R. Noble and J.~R. Torczynski.
	\newblock A lattice-boltzmann method for partially saturated computational
	cells.
	\newblock {\em International Journal of Modern Physics C}, 09(08):1189–1201,
	Dec 1998.
	\newblock \href {https://doi.org/10.1142/S0129183198001084}
	{\path{doi:10.1142/S0129183198001084}}.
	
	\bibitem{Succi_2001}
	Sauro Succi.
	\newblock {\em The lattice Boltzmann equation: for fluid dynamics and beyond}.
	\newblock Oxford university press, 2001.
	
	\bibitem{Haussmann_Hafen_Raichle_Trunk_Nirschl_Krause_2020}
	Marc Haussmann, Nicolas Hafen, Florian Raichle, Robin Trunk, Hermann Nirschl,
	and Mathias~J. Krause.
	\newblock Galilean invariance study on different lattice boltzmann
	fluid–solid interface approaches for vortex-induced vibrations.
	\newblock {\em Computers \& Mathematics with Applications}, 80:671–691, Sep
	2020.
	\newblock \href {https://doi.org/10.1016/j.camwa.2020.04.022}
	{\path{doi:10.1016/j.camwa.2020.04.022}}.
	
	\bibitem{Rettinger_Rüde_2017}
	C.~Rettinger and U.~Rüde.
	\newblock A comparative study of fluid-particle coupling methods for fully
	resolved lattice boltzmann simulations.
	\newblock {\em Computers \& Fluids}, 154:74–89, Sep 2017.
	\newblock \href {https://doi.org/10.1016/j.compfluid.2017.05.033}
	{\path{doi:10.1016/j.compfluid.2017.05.033}}.
	
	\bibitem{Krause_Klemens_Henn_Trunk_Nirschl_2017}
	Mathias~J. Krause, Fabian Klemens, Thomas Henn, Robin Trunk, and Hermann
	Nirschl.
	\newblock Particle flow simulations with homogenised lattice boltzmann methods.
	\newblock {\em Particuology}, 34:1–13, Oct 2017.
	\newblock \href {https://doi.org/10.1016/j.partic.2016.11.001}
	{\path{doi:10.1016/j.partic.2016.11.001}}.
	
	\bibitem{Trunk_Marquardt_Thaeter_Nirschl_Krause_2018}
	Robin Trunk, Jan Marquardt, Gudrun Th\"ater, Hermann Nirschl, and Mathias~J.
	Krause.
	\newblock Towards the simulation of arbitrarily shaped 3D particles using a
	homogenised lattice boltzmann method.
	\newblock {\em Computers {\&} Fluids}, 172:621–631, Aug 2018.
	\newblock \href {https://doi.org/10.1016/j.compfluid.2018.02.027}
	{\path{doi:10.1016/j.compfluid.2018.02.027}}.
	
	\bibitem{Trunk_Weckerle_Hafen_Thaeter_Nirschl_Krause_2021}
	Robin Trunk, Timo Weckerle, Nicolas Hafen, Gudrun Thäter, Hermann Nirschl, and
	Mathias~J. Krause.
	\newblock Revisiting the homogenized lattice boltzmann method with applications
	on particulate flows.
	\newblock {\em Computation}, 9(22):11, Feb 2021.
	\newblock \href {https://doi.org/10.3390/computation9020011}
	{\path{doi:10.3390/computation9020011}}.
	
	\bibitem{Trunk_Bretl_Thaeter_Nirschl_Dorn_Krause_2021}
	Robin Trunk, Colin Bretl, Gudrun Th\"ater, Hermann Nirschl, Márcio Dorn, and
	Mathias~J. Krause.
	\newblock A study on shape-dependent settling of single particles with equal
	volume using surface resolved simulations.
	\newblock {\em Computation}, 9(44):40, Apr 2021.
	\newblock \href {https://doi.org/10.3390/computation9040040}
	{\path{doi:10.3390/computation9040040}}.
	
	\bibitem{Hafen_Dittler_Krause_2022}
	Nicolas Hafen, Achim Dittler, and Mathias~J. Krause.
	\newblock Simulation of particulate matter structure detachment from surfaces
	of wall-flow filters applying lattice boltzmann methods.
	\newblock {\em Computers \& Fluids}, 239:105381, May 2022.
	\newblock \href {https://doi.org/10.1016/j.compfluid.2022.105381}
	{\path{doi:10.1016/j.compfluid.2022.105381}}.
	
	\bibitem{Hafen_2023}
	Nicolas Hafen, Jan~E. Marquardt, Achim Dittler, and Mathias~J. Krause.
	\newblock Simulation of particulate matter structure detachment from surfaces
	of wall-flow filters for elevated velocities applying lattice boltzmann
	methods.
	\newblock {\em Fluids}, 8(3), 2023.
	\newblock \href {https://doi.org/10.3390/fluids8030099}
	{\path{doi:10.3390/fluids8030099}}.
	
	\bibitem{Hafen_Marquardt_Dittler_Krause_2023}
	Nicolas Hafen, Jan~E. Marquardt, Achim Dittler, and Mathias~J. Krause.
	\newblock Simulation of dynamic rearrangement events in wall-flow filters
	applying lattice boltzmann methods.
	\newblock {\em Fluids}, 8(77):213, Jul 2023.
	\newblock \href {https://doi.org/10.3390/fluids8070213}
	{\path{doi:10.3390/fluids8070213}}.
	
	\bibitem{Hafen2023c}
	Nicolas Hafen, Jan~E. Marquardt, Achim Dittler, and Mathias~J. Krause.
	\newblock Simulation of {Dynamic} {Rearrangement} {Events} in {Wall}-{Flow}
	{Filters} {Applying} {Lattice} {Boltzmann} {Methods}.
	\newblock {\em Fluids}, 8(7):213, 2023.
	\newblock \href {https://doi.org/10.3390/fluids8070213}
	{\path{doi:10.3390/fluids8070213}}.
	
	\bibitem{Li2019}
	Xingang Li, Fangzhou Wang, Duo Zhang, Sai Gu, and Xin Gao.
	\newblock Fluid-solid interaction simulation for particles and walls of
	arbitrary polygonal shapes with a coupled lbm-imb-dem method.
	\newblock {\em Powder Technology}, 356:177--192, November 2019.
	\newblock \href {https://doi.org/10.1016/j.powtec.2019.08.006}
	{\path{doi:10.1016/j.powtec.2019.08.006}}.
	
	\bibitem{Marquardt_Römer_Nirschl_Krause_2022}
	Jan~E. Marquardt, Ulrich~J. Römer, Hermann Nirschl, and Mathias~J. Krause.
	\newblock A discrete contact model for complex arbitrary-shaped convex
	geometries.
	\newblock {\em Particuology}, 2022.
	\newblock \href {https://doi.org/10.1016/j.partic.2022.12.005}
	{\path{doi:10.1016/j.partic.2022.12.005}}.
	
	\bibitem{Karimnejad2022}
	S.~Karimnejad, A.~Amiri Delouei, H.~Başağaoğlu, M.~Nazari, M.~Shahmardan,
	G.~Falcucci, M.~Lauricella, and S.~Succi.
	\newblock A review on contact and collision methods for multi-body hydrodynamic
	problems in complex flows.
	\newblock {\em Communications in Computational Physics}, 32(4):899--950, June
	2022.
	\newblock \href {https://doi.org/10.4208/cicp.re-2022-0041}
	{\path{doi:10.4208/cicp.re-2022-0041}}.
	
	\bibitem{Krueger2016}
	Timm Kr{\"u}ger, Halim Kusumaatmaja, Alexandr Kuzmin, Orest Shardt, Goncalo
	Silva, and Erlend~Magnus Viggen.
	\newblock {\em The Lattice Boltzmann Method}.
	\newblock Graduate Texts in Physics. Springer, 2017.
	\newblock \href {https://doi.org/10.1007/978-3-319-44649-3}
	{\path{doi:10.1007/978-3-319-44649-3}}.
	
	\bibitem{sukop2006LatticeBoltzmannModeling}
	Michael~C. Sukop and Daniel~T. Thorne.
	\newblock {\em Lattice Boltzmann Modeling: An introduction for geoscientists
		and Engineers}.
	\newblock Springer, 2., corrected print edition, 2006.
	\newblock \href {https://doi.org/10.1007/978-3-540-27982-2}
	{\path{doi:10.1007/978-3-540-27982-2}}.
	
	\bibitem{Bhatnagar_Gross_Krook_1954}
	P.~L. Bhatnagar, E.~P. Gross, and M.~Krook.
	\newblock A model for collision processes in gases. I. small amplitude
	processes in charged and neutral one-component systems.
	\newblock {\em Physical Review}, 94(3):511–525, May 1954.
	\newblock \href {https://doi.org/10.1103/PhysRev.94.511}
	{\path{doi:10.1103/PhysRev.94.511}}.
	
	\bibitem{Qian1992}
	Y.~H Qian, D~D’Humières, and P~Lallemand.
	\newblock Lattice BGK models for navier-stokes equation.
	\newblock {\em Europhysics Letters (EPL)}, 17(6):479--484, February 1992.
	\newblock \href {https://doi.org/10.1209/0295-5075/17/6/001}
	{\path{doi:10.1209/0295-5075/17/6/001}}.
	
	\bibitem{olb16}
	Adrian Kummerländer, Sam Avis, Halim Kusumaatmaja, Fedor Bukreev, Michael
	Crocoll, Davide Dapelo, Nicolas Hafen, Shota Ito, Julius Jeßberger, Jan~E.
	Marquardt, Johanna Mödl, Tim Pertzel, František Prinz, Florian Raichle,
	Maximilian Schecher, Stephan Simonis, Dennis Teutscher, and Mathias~J.
	Krause.
	\newblock {OpenLB Release 1.6: Open Source Lattice Boltzmann Code}, April 2023.
	\newblock \href {https://doi.org/10.5281/zenodo.7773497}
	{\path{doi:10.5281/zenodo.7773497}}.
	
	\bibitem{Krause2020}
	Mathias~J. Krause, Adrian Kummerländer, Samuel~J. Avis, Halim Kusumaatmaja,
	Davide Dapelo, Fabian Klemens, Maximilian Gaedtke, Nicolas Hafen, Albert
	Mink, Robin Trunk, Jan~E. Marquardt, Marie-Luise Maier, Marc Haussmann, and
	Stephan Simonis.
	\newblock {OpenLB}—{Open} source lattice {Boltzmann} code.
	\newblock {\em Computers \& Mathematics with Applications}, 81:258--288, 2021.
	\newblock \href {https://doi.org/10.1016/j.camwa.2020.04.033}
	{\path{doi:10.1016/j.camwa.2020.04.033}}.
	
	\bibitem{kupershtokh2009EquationsStateLattice}
	A.~Kupershtokh, D.~Medvedev, and D.~Karpov.
	\newblock On equations of state in a lattice {{Boltzmann}} method.
	\newblock {\em Computers \& Mathematics with Applications}, 58:965--974, 2009.
	\newblock \href {https://doi.org/10.1016/j.camwa.2009.02.024}
	{\path{doi:10.1016/j.camwa.2009.02.024}}.
	
	\bibitem{wen2014GalileanInvariantFluid}
	Binghai Wen, Chaoying Zhang, Yusong Tu, Chunlei Wang, and Haiping Fang.
	\newblock Galilean invariant fluid–solid interfacial dynamics in lattice
	{Boltzmann} simulations.
	\newblock {\em Journal of Computational Physics}, 266:161--170, 2014.
	\newblock \href {https://doi.org/10.1016/j.jcp.2014.02.018}
	{\path{doi:10.1016/j.jcp.2014.02.018}}.
	
	\bibitem{Henn_Thäter_Dörfler_Nirschl_Krause_2016}
	Thomas Henn, Gudrun Thäter, Willy Dörfler, Hermann Nirschl, and Mathias~J.
	Krause.
	\newblock Parallel dilute particulate flow simulations in the human nasal
	cavity.
	\newblock {\em Computers \& Fluids}, 124:197–207, Jan 2016.
	\newblock \href {https://doi.org/10.1016/j.compfluid.2015.08.002}
	{\path{doi:10.1016/j.compfluid.2015.08.002}}.
	
	\bibitem{Yin_Koch_2007}
	Xiaolong Yin and Donald~L. Koch.
	\newblock Hindered settling velocity and microstructure in suspensions of solid
	spheres with moderate {R}eynolds numbers.
	\newblock {\em Physics of Fluids}, 19(9):093302, Sep 2007.
	\newblock \href {https://doi.org/10.1063/1.2764109}
	{\path{doi:10.1063/1.2764109}}.
	
	\bibitem{schiller1933uber}
	L~Schiller and A~Neumann.
	\newblock Über die grundlegenden {B}erechnungen bei der
	{S}chwerkraftaufbereitung.
	\newblock {\em Z. Vereines Deutscher Ingenieure}, 77:318--321, 1933.
	
	\bibitem{Ergun_1952}
	Sabri Ergun.
	\newblock Fluid flow through packed columns.
	\newblock {\em Chemical Engineering Progress}, 48(2):89–94, 1952.
	
	\bibitem{Ginzbourg1996}
	I.~Ginzbourg and D.~d’Humières.
	\newblock Local second-order boundary methods for lattice boltzmann models.
	\newblock {\em Journal of Statistical Physics}, 84(5–6):927--971, September
	1996.
	\newblock \href {https://doi.org/10.1007/bf02174124}
	{\path{doi:10.1007/bf02174124}}.
	
\end{thebibliography}

\end{document}